\documentclass[aps,prl,twocolumn, titlepage,showpacs]{revtex4}

\usepackage{graphicx}

\usepackage{dcolumn}

\usepackage{bm}

\begin{document}

\title{Identifying anomalous diffusion and melting in dusty plasmas}

\author{Yan Feng}
\email{yan-feng@uiowa.edu}
\author{J. Goree}
\author{Bin Liu}
\affiliation{Department of Physics and Astronomy, The University
of Iowa, Iowa City, Iowa 52242}

\date{\today}

\begin{abstract}

Anomalous diffusion in liquids and the solid-liquid phase
transition (melting) are studied in two-dimensional Yukawa
systems. The self-intermediate scattering function (self-ISF),
calculated from simulation data, exhibits a temporal decay, or
relaxation, with a characteristic relaxation time. This decay is
found to be useful for distinguishing normal and anomalous
diffusion in a liquid, and for identifying the solid-liquid phase
transition. For liquids, a scaling of the relaxation time with
length scale is found. For the solid-liquid phase transition, the
shape of the self-ISF curve is found to be a sensitive indicator
of phase. Friction has a significant effect on the timing of
relaxation, but not the melting point.

\end{abstract}

\narrowtext

\maketitle

\section {I.~INTRODUCTION}

Dusty plasma is partially ionized gas containing micro-sized
particles of solid matter \cite{Feng:08}. In a plasma
\cite{Feng:08}, the sheath above a lower electrode has electric
fields that can levitate and confine highly charged particles, so
that they are suspended. When only a single layer is suspended,
the interaction between dust particles is a repulsive Yukawa
potential \cite{Konopka:00}. Video microscopy allows imaging this
two-dimensional (2D) suspension at an atomistic scale, so that we
can track particles and measure their individual positions and
velocities in each video frame, yielding the same kind of data as
the molecular dynamics (MD) simulations, reported here.

Particles self-organize in a crystal-like triangular lattice with
hexagonal symmetry due to strong interparticle interaction. In
this strongly-coupled plasma, the Coulomb interaction with nearest
neighbors is so strong that particles do not easily move past one
another \cite{Ichimaru:82}. In many experiments, the particles
occupy only a single horizontal layer, and are often described as
2D experiments \cite{Feng:08, Nosenko:06, Nunomura:05, Ivlev:03,
Samsonov:99, Ratynskaia:06}. Dusty plasma is a driven-dissipative
system \cite{Feng:08}, and its kinetic energy is determined by the
balance of the energy input and dissipation. As the driven energy
input increases, the lattice becomes disordered in a solid-liquid
phase transition \cite{Thomas:96, Melzer:96, Quinn:01, Nosenko:06,
Knapek:07, Feng:08, Sheridan:08}.

Random particle motion in dusty plasmas can be divided into
several stages. Ballistic motion \cite{Liu:08} occurs on a short
time scale $< \omega_{pd}^{-1}$, while caging oscillations
\cite{Io:09} happen on a typical time scale of
$10~\omega_{pd}^{-1}$. Here $\omega_{pd}$ is the nominal plasma
frequency \cite{Kalman:04}. At later times, particles can escape
their cages and diffuse. A current research topic that has not
been resolved is whether this long-time random motion exhibits
normal diffusion or anomalous diffusion. Experiments \cite{Liu:08,
Liu:08_2} and simulations \cite{Liu:07, Hou:09, Donko:09} with
these 2D systems have indicated superdiffusion, where the
mean-square-displacement (MSD) increases with the time more
rapidly than linear scaling, but other simulations suggest that
random motion may be normal diffusion at sufficiently long times
\cite{Ott:09}. Identifications of normal diffusion and anomalous
diffusion (e.g. superdiffusion here) have typically made use of
dynamical measures: MSD time series, the probability distribution
function (PDF) for particle displacements, and the velocity
autocorrelation function (VACF)~\cite{Liu:07, Donko:09, Hou:09,
Ott:09}.

Data analysis methods used to study the solid-liquid phase
transitions can be grouped in two categories, static and dynamic.
Here, we term a method static if the input data can be a single
snapshot of particle positions, but dynamic if it requires a
series of positions for each particle. According to this
classification, most attempts to identify phase transitions in
dusty plasma experiments and MD simulations have employed only
static structural measures, such as Voronoi diagrams and
correlation functions for particle positions and angular
orientation \cite{Thomas:96, Melzer:96, Quinn:01, Hartmann:05,
Nosenko:06, Knapek:07, Feng:08, Sheridan:08}. Sheridan used a
static method of applying the empirical Lindemann
criterion~\cite{Lindemann:10, Sheridan:08}. A dynamical method of
identifying melting has been developed theoretically, using
empirical criteria based on the long-time and short-time
self-diffusion coefficients~\cite{Lowen:93, Lowen:96}; this has
been demonstrated to be useful for simulations of both
2D~\cite{Lowen:96} and 3D~\cite{Lowen:93} systems.

Here we carry out dynamical analysis using the intermediate
scattering function to characterize random motion. We study two
physical processes: anomalous diffusion in liquids and the
solid-liquid phase transition.

In Sec.~II, we will review the intermediate scattering function
briefly. In Sec.~III, we will introduce our two MD Yukawa
simulation methods: Langevin and frictionless. They will model a
2D dusty plasma. In Sec.~IV, we will present results for the two
physical processes: normal and anomalous diffusion in liquids, and
the melting phase transition.

\section{II. SELF-INTERMEDIATE SCATTERING FUNCTION}

The intermediate scattering function (ISF) \cite{Pusy:91}, which
has been used widely in other fields, is also called the
density-density correlation function \cite{Kambayashi:87}. The ISF
is defined in terms of the particle trajectories:
\begin{equation}\label{ISF}
{F(\mathbf{k},t)=\frac{1}{N} \sum_i \sum_j \langle \rm{exp}
(-i\mathbf{k}\cdot [\mathbf{r}_i(t)-\mathbf{r}_j(0)])\rangle .}
\end{equation}
Here, $\mathbf{r}_i(t)$ is the trajectory of the $i$th particle in
the system consisting of $N$ particles. The Fourier transform
variable $\mathbf{k}$ is usually called a wave number, although no
waves are studied using this method. Equation~(\ref{ISF}) makes
use of an ensemble average $\langle \cdot\cdot\cdot \rangle$,
which in practice is done by averaging for various initial
starting times in place of $t=0$.

Calculating the ISF directly from Eq.~(\ref{ISF}) is a dynamical
analysis method because it requires as its input data a time
series of positions for each particle. This is the method we will
use, as has been done previously in experiments with granular
materials \cite{Dauchot:05, Reis:07} and in MD simulations
\cite{Kob:97, Meyer:10, Abete:07} of systems other than dusty
plasmas.

Besides starting from measurements of particle trajectories, other
experimental methods of obtaining the ISF have been devised for
colloids \cite{Megen:91}, supercooled liquids
\cite{Bellissent-Funel:00}, and polymer nanocomposites
\cite{Srivastava:09}. In these experiments, the ISF was determined
from data produced by dynamic light scattering \cite{Megen:91},
x-ray photon correlation spectroscopy \cite{Lu:08, Srivastava:09},
or neutron spin echo spectroscopy \cite{Hansen:86,
Bellissent-Funel:00}. A comparable experiment with a dusty plasma
was reported by Khodataev {\it et al.} \cite{Khodataev:98}, using
photon correlation spectroscopy to yield a function of wave number
and time.

It is useful to compare the ISF, which is a time series, with two
related functions. First, the dynamic structure
factor~\cite{Hansen:86} is the Fourier transform of the ISF; it is
defined as
$S(\mathbf{k},\omega)=\int^\infty_{-\infty}F(\mathbf{k},t)~{\rm
exp}(i\omega t)~dt/2\pi$. This dynamic structure factor has been
used for studying 2D Yukawa liquids in both
theory~\cite{Murillo:03} and simulations~\cite{Hou:06}. Second,
phonon spectra, which are widely used for experimental and
simulation data, have two parts, for longitudinal and transverse
waves. The longitudinal phonon spectrum is related to the dynamic
structure factor~\cite{Donko:08}, although phonon spectra are more
commonly computed differently, as a Fourier transform of the
particle current~\cite{Liu:09}.

The ISF is composed of two parts \cite{Hansen:86, Megen:98},
$F(\mathbf{k},t)=F_s(\mathbf{k},t)+F_c(\mathbf{k},t)$. The most
commonly used part is $F_s(\mathbf{k},t)$, which is often called
the {\it incoherent} part, or the {\it self-ISF}:
\begin{equation}\label{selfISF}
{F_s(\mathbf{k},t)=\frac{1}{N} \sum_i \langle \rm{exp}
(-i\mathbf{k}\cdot [\mathbf{r}_i(t)-\mathbf{r}_i(0)])\rangle .}
\end{equation}
The less commonly used part is $F_c(\mathbf{k},t)$, which is
called the {\it coherent} part:
\begin{equation}\label{c-ISF}
{F_c(\mathbf{k},t)=\frac{1}{N} \sum_{i\neq j} \sum_j \langle
\rm{exp} (-i\mathbf{k}\cdot
[\mathbf{r}_i(t)-\mathbf{r}_j(0)])\rangle .}
\end{equation}

The self-ISF, $F_s(\mathbf{k},t)$, is a measure of single-particle
dynamics as a function of time. This makes it comparable to the
MSD and PDF, which are also computed from the trajectories of
individual particles recorded for a long time. Thus, the self-ISF
can be used to study some of the same physical phenomena as MSD
and PDF, such as random motion and the related idea of relaxation
\cite{Reis:07}. If random motion consists of normal diffusion, as
for example with Brownian motion with a diffusion coefficient $D$,
then \cite{Megen:98, Dauchot:05, Reis:07}
\begin{equation}\label{selfISF-diffusion}
{F_s(\mathbf{k},t) \simeq {\rm exp}(-Dk^2t).}
\end{equation}
In Sec.~IV~A~2, we will generalize Eq.~(\ref{selfISF-diffusion})
for the case of anomalous diffusion, such as superdiffusion.

The self-ISF, Eq.~(\ref{selfISF}), is often used by itself,
without reporting the coherent part, Eq.~(\ref{c-ISF}). This is a
common practice with particle trajectory data from experiments
\cite{Dauchot:05, Reis:07} and MD simulations \cite{Meyer:10,
Kob:97, Abete:07} for various physical systems. Here, we will also
use only the self-ISF for our Yukawa simulations of dusty plasmas.

A graph of the self-ISF typically reveals two stages of random
motion. We illustrate this in Fig.~1 with a sketch of
$F_s(\mathbf{k},t)$ for a normal liquid (not supercooled). The
decay of the $F_s(\mathbf{k},t)$ curve is the signature of what is
often called relaxation. Caging motion is indicated at short
times. This early part of the curve is sometimes termed the fast
$\beta$ relaxation. Diffusion is indicated at long times, as
particles gradually escape their cages \cite{Reis:07}. In this
later part of the curve, sometimes termed $\alpha$ relaxation,
$F_s(\mathbf{k},t)$ gradually decays toward zero. Here, we will
use the self-ISF two ways. We will inspect the shape of the
self-ISF decay, and sometimes fit it to a stretched exponential
\cite{Ediger:96}:
\begin{equation}\label{s-exponential}
{F_s(\mathbf{k},t)={\rm exp}[-(t/\tau(k))^{\beta(k)}],}
\end{equation}
where $\tau(k)$ is a relaxation time. (Note that this use of the
symbol $\beta$ has no relation to the $\beta$ relaxation.)

\section {III.~SIMULATION}

\subsection{A. Parameters}
Equilibrium Yukawa systems are characterized by two dimensionless
parameters: the coupling parameter $\Gamma$ and the screening
parameter $\kappa$~\cite{Ohta:00, Sanbonmatsu:01}. Here,
$\Gamma=Q^2/(4\pi\epsilon_0ak_BT)$ and $\kappa\equiv a/\lambda_D$,
where $Q$ is the particle charge, $T$ is the particle kinetic
temperature, $\lambda_D$ is the screening length, $a\equiv
(n\pi)^{-1}$ is the Wigner-Seitz radius~\cite{Kalman:04}, and $n$
is the areal number density. Another characteristic length is the
lattice constant $b$ for a defect-free crystal, which is $b =
1.9046~a$ for a 2D triangular lattice.

Our two simulation methods are the same in many respects. Both
simulation methods use a binary interparticle interaction with a
Yukawa pair potential,
\begin{equation}\label{Yukawa}
\phi_{i,j}=Q^2(4\pi\epsilon_0r_{i,j})^{-1}{\rm
exp}(-r_{i,j}/\lambda_D),
\end{equation}
where $r_{i,j}$ is the distance between the $i$th and $j$th
particles. In both simulations, particles are only allowed to move
in a single 2D plane. Conditions remained steady during each
simulation run. For both simulations, the parameters we used were
$N = 16~384$ particles in a rectangular box with periodic boundary
conditions. The box had sides $137.5\thinspace
b~\times~119.1\thinspace b$. We truncated the Yukawa potential at
radii beyond $12~b$ \cite{Feng:08_2}. The integration time step
was $0.037~\omega_{pd}^{-1}$, and simulation data were recorded
for a time duration of $1777~\omega_{pd}^{-1}$ in after the system
reached its steady state. Other simulation details are presented
in \cite{Feng:08_2, Liu:07}. We report results with distances
normalized by $b$, while time (and frictional damping rate $\nu$)
are normalized using the nominal plasma frequency
$\omega_{pd}=(Q^2/2\pi\epsilon_0ma^3)^{1/2}$~\cite{Kalman:04},
where $m$ is the particle mass.

We will next review the two simulation methods. They differ mainly
in the equations of motion that are solved.

\subsection{B. Langevin MD simulations}

Our Langevin MD simulations take into account the dissipation due
to frictional gas damping. The Langevin equation \cite{Liu:07,
Donko:09, Hou:09, Ott:09, Vaulina:09, Feng:08_2} of motion for
each particle is
\begin{equation}\label{LDmotion}
m\ddot{\mathbf{r}}_{i}=-\nabla \sum \phi_{ij}-\nu
m\dot{\mathbf{r}}_{i}+\zeta_{i}(t).
\end{equation}
Trajectories $\mathbf{r}_i(t)$ are found by integrating
Eq.~(\ref{LDmotion}) for all particles. Terms on the right-hand
side include a frictional drag $\nu m\dot{\mathbf{r}}_{i}$ and a
random force $\zeta_{i}(t)$. Note that we retain the inertial term
on the left-hand-side in Eq.~(\ref{LDmotion}), unlike some
Brownian-dynamics simulations of overdamped colloidal
suspensions~\cite{Lowen:92}, where it is set to zero.

Our Langevin simulations mimic 2D dusty plasma
experiments~\cite{Feng:08}, but the driven-dissipation mechanism
is only an approximation of the processes in
experiments~\cite{Feng:08_2}. In our Langevin simulation, the
heating and friction are explicitly coupled by the
fluctuation-dissipation theorem~\cite{Pathria:1972, Gunsteren:82};
this models collisions with gas atoms that provide both frictional
drag and random kicks. However, besides random kicks from gas
atoms, in dusty plasma there are some additional heating
mechanisms arising from ion flow, particle charge
fluctuations~\cite{Quinn:00, Norman:10}, and sometimes external
laser manipulation \cite{Feng:08, Feng:08_2} that are not
explicitly modeled in our Langevin simulations.

\subsection{C. Frictionless equilibrium MD simulations}

In addition to our Langevin MD simulations which include friction,
to obtain results in the frictionless limit, we also performed
frictionless equilibrium MD simulations \cite{Liu:07}. The
equation of motion is
\begin{equation}\label{MDmotion}
m\ddot{\mathbf{r}}_{i}=-\nabla \sum \phi_{ij},
\end{equation}
which we integrate for all particles. A Nos\'e-Hoover thermostat
is applied to maintain a desired temperature \cite{Liu:07}.

This MD simulation method describes a frictionless atomic system.
The particles collide among themselves, without any interaction
with gas or other external influences. This method mimics thermal
equilibrium conditions.

There are two parameters we can change in the frictionless MD
simulations: $\Gamma$ and $\kappa$. In the Langevin simulations,
we can also vary $\nu$. Varying $\Gamma$ and $\kappa$ is
equivalent to varying temperature and density, and we will vary
them over a range that allows us to simulate liquids or solids.

Our method is to generate trajectories $\mathbf{r}_i(t)$ for all
particles by integrating Eq.~(\ref{LDmotion}) or (\ref{MDmotion}),
and then to compute the self-ISF using Eq.~(\ref{selfISF}). The
self-ISF is a time series. We repeat its calculation for various
wave numbers, $k$.

\section {IV.~RESULTS AND DISCUSSIONS}

We present results for two physical processes for 2D systems under
steady conditions. First, we evaluate whether random motion in a
2D liquid exhibits normal or anomalous diffusion. Second, we test
whether the self-ISF can serve as a sensitive indicator of the
phase transition between solid and liquid.

In this paper, relaxation refers to the decay of the self-ISF,
$F_s(\mathbf{k},t)$, as shown in Fig.~1. The relaxation rate in
general depends on the scale length, which is parameterized here
by $1/k$. Using the terminology of other users of the self-ISF
\cite{Reichman:05, Reis:07}, the early stage of decay is termed
$\beta$ relaxation; during this early time particles are mainly
trapped within their cages formed by nearest neighbors. The later
stage is termed $\alpha$ relaxation, and this corresponds to
diffusion as particles decage. The term ``decaging'' refers to a
particle's movement so that it is no longer trapped by the
previous nearest neighbors. In a liquid, particles decage much
more rapidly than in a solid, so that this $\beta$ relaxation is
much faster in liquids than solids.

\subsection{A. Anomalous diffusion in 2D liquids}

\subsubsection{1. Results for the self-ISF}

Results from our Langevin simulations are presented in Fig. 2 for
typical liquid conditions far from the phase transition. Curves
are shown as functions of time for various values of $k$. Note the
smooth and gradual decay from unity to zero as time increases.
This decay is the signature of relaxation. Here, the decay
develops without any plateau. The lack of a plateau in the
time-variation of the self-ISF is similar to what is seen in
granular flows \cite{Dauchot:05, Reis:07}, but different from what
is expected for supercooled liquids and glasses (cf. Fig. 3 of
\cite{Reichman:05}).

To help quantify the relaxation that is observed in Fig.~2, we fit
the time-dependence of the self-ISF to the empirical form
Eq.~(\ref{s-exponential}). Since the relaxation process spans many
decades of time, to perform this fit without biasing results
toward long times, data points were sampled from the simulation
results at time intervals equally spaced on a logarithmic scale.
For a liquid far from the phase transition,
Eq.~(\ref{s-exponential}) fits our simulation data points well
(shown as solid lines in Fig.~2). The two free parameters for the
fit, $\tau(k)$ and $\beta(k)$, help quantify the relaxation
process. We discuss their physical significance below.

\subsubsection{2. Searching for anomalous diffusion}

We develop two tools for identifying anomalous diffusion. Previous
investigators have usually used the MSD, looking for a scaling
with time that differs from the MSD $\propto t$ scaling expected
for normal diffusion \cite{Hou:09, Ott:09, Liu:08, Liu:08_2,
Liu:07}. For 2D systems, data are typically fit to the form
\begin{equation}\label{MSD}
{\langle r^2(t) \rangle = 4D~t^\alpha,}
\end{equation}
where $\alpha = 1$ is the case of normal diffusion, $\alpha > 1$
is superdiffusion, and $\alpha < 1$ is subdiffusion. (This use of
the symbol $\alpha$ has no relation to the $\alpha$ relaxation
mentioned above.) Here we introduce two other tools that are also
based on how random motion develops with time: the scaling with
$\tau$ vs. $k$, and the value of $\beta$. Recall that $\tau$ and
$\beta$ are the fitting parameters for Eq.~(\ref{s-exponential}).

Our first new tool is the power-law scaling of the fitting
parameter $\tau$ as compared to $k$. To do this, we must first
generalize Eq.~(\ref{selfISF-diffusion}) to allow for anomalous
diffusion. Starting from Eq.~(\ref{selfISF}), previous authors
\cite{Megen:98, Nijboer:66} have demonstrated that
\begin{equation}\label{selfISF-diffusion2}
{F_s(\mathbf{k},t) \simeq {\rm exp}(-\frac{k^2 \langle r^2(t)
\rangle}{4}),}
\end{equation}
where we have substituted $4$ in place of $6$ in the denominator
for two dimensions instead of three. Next, we substitute
Eq.~(\ref{MSD}) in Eq.~(\ref{selfISF-diffusion2}),  yielding
\begin{equation}\label{selfISF-diffusion-general}
{F_s(\mathbf{k},t) \simeq {\rm exp}(- k^2 D t^\alpha) = {\rm
exp}(- D (k^{2/\alpha}t)^\alpha).}
\end{equation}
Examining the argument on the right-hand-side reveals the scaling
\begin{equation}\label{scaling}
{\tau \propto k^{-2/\alpha}.}
\end{equation}
In the case of normal diffusion, $\alpha = 1$, the scaling is
$\tau \propto k^{-2}$, as previous authors have noted
\cite{Megen:98, Dauchot:05, Reis:07}. Here, we note that the
superdiffusion case $\alpha > 1$ has $\tau$ varying with a lesser
power. Thus, the signature of superdiffusion will be a slope
weaker than $-2$ when $\tau$ is plotted vs. $k$ using log-log
axes. Similarly, a slope stronger than $-2$ in the same $\tau - k$
plot is the signature of subdiffusion.

Our second new tool is the fitting parameter $\beta$ for the
self-ISF. Comparing Eq.~(\ref{s-exponential}) and
Eq.~(\ref{selfISF-diffusion-general}), we see that the value of
$\beta$ is essentially the same as $\alpha$. The only difference
is that when using actual data, the value of $\beta$ is generated
by a fit, while $\alpha$ is generated by examining a log-log plot.
The community of scientists who use the self-ISF traditionally
uses $\beta$, although until now it has not been used as an
indicator of superdiffusion. Scientists who use MSD to
characterize superdiffusion, on the other hand, traditionally use
$\alpha$.

The two new diagnostic tools we described above may be useful for
experiments in other fields where measurements allow one to obtain
the self-ISF. While methods exist to distinguish anomalous
diffusion from normal diffusion (using MSD, PDF, or VACF) if
particle tracking is feasible, as in dusty plasma
experiments~\cite{Liu:08} and MD simulations \cite{Liu:07, Hou:09,
Donko:09, Ott:09}, other methods are needed when tracking is
impossible. For example in a simple liquid, the motion of
individual molecules cannot be tracked, but the self-ISF can be
obtained using dynamic light scattering~\cite{Megen:98} or some
other spectroscopic methods \cite{Lu:08, Srivastava:09, Hansen:86,
Bellissent-Funel:00}. Thus, the two diagnostic tools presented
above could find an application in such experiments.

\subsubsection{3. Results for fitting the self-ISF}

For the conditions of a liquid far from the phase transition, we
use our two tools (scaling of $\tau$ vs. $k$ and the value of
$\beta$) to test for anomalous diffusion. We use our two MD
simulations, Langevin and frictionless. Ott and Bonitz
\cite{Ott:09} previously varied the values of friction $\nu$ and
observation time over wide ranges, and using the MSD method
prepared a diagram showing the conditions that favor normal
diffusion or superdiffusion. This diagram, Fig.~3 of
\cite{Ott:09}, predicts that the value of $\nu$ which we use in
our Langevin simulation will yield normal diffusion, while the
$\nu = 0$ case of our frictionless simulation will yield
superdiffusion over any reasonable observation time. Here we test
whether our two cases, analyzed using our two new tools, yield the
same conclusion as in \cite{Ott:09}.

For the $\tau$ vs. $k$ scaling, in Fig. 3(a) we find the scaling
$\tau \propto k^{-2}$ for the Langevin MD simulations, and the
scaling of $\tau \propto k^{-\gamma}$ ($\gamma<2$) for the
frictionless MD simulations. In other words, random motion is
diffusive for our frictional (Langevin) case, but superdiffusive
for our frictionless case. This is in quantitative agreement with
Fig.~3 of \cite{Ott:09}, prepared using MSD curves, which
demonstrated that friciton can inhibit superdiffusion
\cite{Ott:09, Hou:09}.

For $\beta$, in Fig. 3(b) we find values near unity for our
Langevin simulation, but a value definitely $>1$ for the
frictionless simulation, for moderate values of $k$. This is again
consistent with the conclusion of diffusive motion for our
frictional (Langevin) case, but superdiffusive motion for the
frictionless case. At extremely small or large values of $k$,
however, $\beta$ can be different. Figure~3(b) reveals an overall
trend for $\beta$ to increase with $k$, especially at extremely
small or large values of $k$. Previous authors \cite{Dauchot:05,
Reis:07} have identified dynamic heterogeneities as the cause for
$\beta < 1$ for very small $k$, i.e., very large length scales.
For large $k$, previous authors have not reported enhanced values
of $\beta$ like those we see in Fig. ~3(b). One possible
interpretation of our large $k$ observation is that, at these
short length scales, the self-ISF is affected more by caging
motion than by random walks associated with decaging.

To summarize, we find that the relaxation of the self-ISF is a
sensitive indicator to distinguish anomalous diffusion from normal
diffusion. The indication can be made using either of the two
fitting parameters, $\tau$ and $\beta$.

\subsection{B. Solid-liquid phase transition}
Simulation studies of the solid-liquid phase transition, which is
also called an order-disorder transition or melting, are generally
done using measures of static structural order, such as defects or
correlation functions of particle position or bond orientation
\cite{Thomas:96, Melzer:96, Quinn:01, Hartmann:05, Nosenko:06,
Knapek:07, Feng:08, Sheridan:08}. Dynamical measures can provide
additional information that can be helpful for identifying a
phase. Temperature is kind of dynamical measure for random motion,
and so is the self-ISF.

Our first goal, for phase transitions, is to test the use of the
self-ISF as an indicator of the phase transition. We perform tests
that indicate that it is sensitive in distinguishing solids and
liquids near the phase transition. This development is useful
because the self-ISF is based on dynamics rather than structure.
Our second goal is to determine what role friction plays in the
phase transition. We will vary temperature and density, using the
normalized quantities $\Gamma$ and $\kappa$, and we will also vary
the friction $\nu$, to determine whether the self-ISF is sensitive
to phases, and what role friction plays.

\subsubsection{1. Dependence on $\Gamma$}
We find that the self-ISF time series undergoes a sudden change at
the melting point. This is seen for the self-ISF in Fig.~4 for our
Langevin simulation. Here, we have chosen to present results for a
small wave number $k=2\pi/b$, which is the wave number
corresponding to a lattice constant so that the self-ISF indicates
dynamics at the length scale of nearest neighbors. As we varied
$\Gamma$ in Fig.~4, we held $\kappa=1.2$ and
$\nu/\omega_{pd}=0.027$ as constants.

We note two features of the self-ISF curves in Fig.~4 that are
different on either side of this sudden change. First, the gap
between curves is much wider for low temperature (high $\Gamma$)
conditions in the upper right of the figure as compared to the
high temperature conditions in the lower left. We varied $\Gamma$
in small steps near $\Gamma=200$, where we find the sudden change
in the gaps between the self-ISF curves. Second, we identify
different shapes for the decay of the self-ISF for low and high
temperatures. For the high temperature (low $\Gamma$) conditions
expected for liquids we found, in Sec.~IV~A, that the self-ISF
decays according to the empirical law Eq.~(\ref{s-exponential}),
but for low temperatures we found that Eq.~(\ref{s-exponential})
does not come even close to the shape of the curves in the upper
right of Fig.~4.

Comparing to previous simulations that used structural measures,
we can confirm that the sudden change in the self-ISF curve
corresponds to the phase transition. Using a measure of local
orientation order that exhibited a large jump at the phase
transition, a phase transition curve for $\Gamma$ vs. $\kappa$ was
reported,  Fig.~6 in \cite{Hartmann:05}. Interpolating their
results, we find that the phase transition occurs at about $\Gamma
=200$ for $\kappa=1.2$, for a 2D Yukawa system modeled with a
frictionless MD simulation. This result is consistent with the
sudden change that we observed in our self-ISF curves for $\kappa
= 1.2$ in Fig.~4: first, the curve's shape is changed; second, the
curves are narrowly spaced for liquids ($\Gamma<200$) and widely
spaced for solids ($\Gamma>200$). (One difference in the
simulations of \cite{Hartmann:05} and ours is our use of friction
comparable to values in 2D dusty plasma experiments.  We will
explore the role of friction in Sec.~IV~B~4.)

Thus, we conclude that the self-ISF curve is very sensitive to
phase. It shows promise to become a reliable indicator of the
phase transition, although further tests, for different physical
parameters would be needed to confirm its reliability.

An attraction of this method is that the self-ISF is a dynamical
rather than structural measure. Combining both dynamical and
structural measures can be useful in distinguishing phases. For
example, supercooled liquids have the structure of normal liquids,
so that a dynamical measure is needed to distinguish the two. We
note that other methods of using dynamical measures for such
purposes include the L\"owen criterion~\cite{Lowen:93, Lowen:96},
which involves a comparison of diffusion coefficients computed in
two limiting cases.

\subsubsection{2. Dependence on $\kappa$}
Varying only the density or $\kappa$, we again find the same two
results as for varying $\Gamma$. There is a sudden change in the
gaps between curves, and the curves take a different shape at a
point that we can identify as the phase transition. This is seen
in Fig.~5, where we changed $\kappa$ in our Langevin simulations
with $\Gamma=200$ and $\nu/\omega_{pd}=0.027$. The transition
occurs at $\kappa=1.2$, which is consistent with Fig.~6 in
\cite{Hartmann:05}.

\subsubsection{3. Structure relaxation time}
We investigate how rapidly disorder develops on the length scale
of a cage, i.e., the interparticle spacing $b$. We will test
whether it occurs with an Arrhenius or Vogel-Fulcher law as in
other complex fluids such as colloids \cite{Kurita:09}, granular
materials \cite{Reis:07}, supercooled liquids and glasses
\cite{Kawasaki:07}. We carry out this investigation by
characterizing a decay time for the self-ISF. We could use the fit
parameter $\tau(k)$ from Eq.~(\ref{s-exponential}), but for
simplicity, rather than fitting the self-ISF, here we will adopt
the practice of other authors of measuring the time required for
the self-ISF to decay by a factor of $1/e$ \cite{Kawasaki:07,
Kurita:09}. This is done only for $k = 2 \pi / b$, corresponding
to the length scale of a cage. Following the terminology used for
example in the literature for colloids \cite{Kurita:09},
supercooled liquids and glasses \cite{Kawasaki:07}, this $1/e$
decay time, denoted here as $\tau_{2 \pi / b}$, is called the
``structure relaxation time.'' It is in principle the same as
$\tau$ if $\beta = 1$.

Results for the structure relaxation time  $\tau_{2 \pi / b}$ for
the curves of Figs.~4-5 are shown in Fig.~6, revealing how
$\tau_{2 \pi / b}$ varies with temperature and number density. As
the normalized temperature $1/\Gamma$ increases, the system melts
and $\tau_{2 \pi / b}$ decreases about one order of magnitude. The
rate at which a structure relaxes, $1/\tau_{2 \pi / b}$, increases
with temperature, which is the same trend, for example, as for
diffusion~\cite{Vaulina:02}. Melting also occurs as $\kappa$ is
increased. Plotting $\tau_{2 \pi / b}$ vs. number density, in the
inset of Fig.~6(b), we note a nearly linear scaling in the liquid
regime.

The scaling we observe for $\tau_{2 \pi / b}$ vs. number density
is different from the Arrhenius and Vogel-Fulcher laws. The
Vogel-Fulcher law, which has been found empirically for other
complex fluids \cite{Kawasaki:07, Kurita:09, Reis:07}, has
$\tau_{2 \pi / b}$ diverging to infinity as the number density
increases toward the phase transition point. Here, $\tau_{2 \pi /
b}$ increases about an order of magnitude near the phase
transition, but it does not diverge.

\subsubsection{4. Dependence on $\nu$}
We find that friction can play an important role in the random
motion of 2D Yukawa liquids, but only at large damping rates. This
result is shown in Fig.~7 where we varied $\nu$ while holding
$\Gamma$ and $\kappa$ constant in a liquid regime. We observe that
the self-ISF curves all lie on top of one another for
$\nu/\omega_{pd} \lesssim 0.03$, but not at higher damping rates
where they decay more slowly as $\nu$ is increased. The curves at
higher damping rates have the same general shape, but are retarded
in time. In other words, friction has a significant effect on the
timing of relaxation, but not the melting point. For the limiting
case of no friction, we also include in Fig.~7 results for our
frictionless equilibrium MD simulation, and these agree with the
Langevin simulation for low friction $\nu/\omega_{pd} \lesssim
0.03$. We note that Vaulina {\it et al.} \cite{Vaulina:02} found
that diffusion is mostly independent of damping rate, in the limit
of low friction, which is similar to our results for relaxation.

We interpret the difference in the $F_s(2\pi/b,t)$ curves at
various damping rates as indicating retardation of diffusion at
high friction levels. At higher damping rate, the random motion of
particles is resisted, then more energy is dissipated locally. As
a result, the collective relaxation, which refers to the
diffusion, will be retarded.

\section {V.~CONCLUSIONS}
In conclusion, we have performed Yukawa MD simulations to study
two physical processes in 2D dusty plasmas: anomalous diffusion in
liquids and melting. For liquids, examining the decay or
relaxation of the self-ISF reveals a scaling of the relaxation
time vs. length scale. This scaling is demonstrated to be useful
for distinguishing normal and anomalous diffusion. The self-ISF is
also demonstrated to be a sensitive indicator of the solid-liquid
phase transition, i.e., melting. Friction has the effect of
retarding relaxation.

This work was supported by NSF and NASA.

\begin{figure}[p]
\caption{\label{ISF-sketch} A typical example of the
self-intermediate scattering function (self-ISF)
$F_s(\mathbf{k},t)$, sketched here to show how it begins at unity
at $t = 0$, and then relaxes gradually to zero as $t \rightarrow
\infty$. Relaxation happens in two steps: early-time caging
motion, and long-time diffusion. The self-ISF curve is often
modeled by Eq.~(\ref{s-exponential}). The case shown here is for a
normal liquid in 2D, and to reveal phenomena at a length scale
corresponding to caging we chose $k = 3.3~a$, where $a$ is the
Wigner-Seitz radius, as defined in Sec.~III.}
\end{figure}

\begin{figure}[p]
\caption{\label{ISF-k} (color online). Time dependence of the
self-ISF for various wave numbers ($k$) for the Langevin molecular
dynamics (MD) simulation in the liquid regime: $\Gamma=200$,
$\kappa=2$ and $\nu/\omega_{pd}=0.027$. The solid lines are the
corresponding fits to Eq.~(\ref{s-exponential}). Quantities are
normalized by parameters, such as the lattice constant $b$, as
defined in Sec.~III~A.}
\end{figure}

\begin{figure}[p]
\caption{\label{fitting} (color online). The fitting parameters,
relaxation time $\tau$ in (a) and exponent $\beta$ in (b), as a
function of $k$. These cross symbols are from fitting the data in
Fig.~2 for the Langevin MD simulation, and the circle symbols are
for the frictionless MD simulation for the same liquid conditions,
$\Gamma=200$, $\kappa=2$. We find that the scaling in (a) and the
values of $\beta$ in (b) are useful as indicators of anomalous
diffusion.}
\end{figure}

\begin{figure}[p]
\caption{\label{ISF-temp} (color online). The self-ISF for the
length scale corresponding to the lattice constant, $k=2\pi/b$.
Results shown are for the Langevin MD simulations with constant
values of $\kappa=1.2$ and $\nu/\omega_{pd}=0.027$. To search for
an indication of a phase transition, $\Gamma$ is varied over a
range spanning both the solid and liquid phases. A phase
transition near $\Gamma=200$ is indicated by a change in the
curve's shape, and by larger gaps between curves for solids,
$\Gamma>200$.}
\end{figure}

\begin{figure}[p]
\caption{\label{ISF-screening} (color online). The self-ISF at
$k=2\pi/b$, as in Fig.~4 but with varying $\kappa$. The
simulations here are all Langevin MD simulations with constant
values of $\Gamma=200$ and $\nu/\omega_{pd}=0.027$. To search for
an indication of a phase transition, $\kappa$ is varied over a
range spanning both the solid and liquid phases. A phase
transition near $\kappa=1.2$ is indicated by a change in the
curve's shape, and the gaps between curves.}
\end{figure}

\begin{figure}[p]
\caption{\label{structure} Structure relaxation time measured from
Figs.~\ref{ISF-temp} and ~\ref{ISF-screening} as a function of (a)
temperature and (b) screening parameter $\kappa$. The structure
relaxation time in our 2D underdamped Yukawa system does not obey
the Arrhenius law or Vogel-Fulcher law. The same data are plotted
in the inset as a function of number density $1/\kappa^2$, for the
liquid regime.}
\end{figure}

\begin{figure}[p]
\caption{\label{ISF-damping} (color online). The role of friction
in relaxation is investigated by computing the self-ISF at
$k=2\pi/b$, for simulations of a liquid. The simulations here
include both Langevin MD simulations and frictionless MD
simulations in a liquid regime, with constant values of
$\Gamma=200$ and $\kappa=3$. Unlike temperature and number
density, changing the friction does not significant change the
shape of the curves. The effect of friction in the self-ISF curves
is a retardation of the decay at higher damping rate.}
\end{figure}

\end{document}